\documentclass[12pt,english]{article}
\usepackage{titlesec}

\titleformat{\section}
  {\centering\normalfont\fontsize{12}{12}\bfseries} 
  {\thesection.}{1em}{\MakeUppercase} 

\titleformat{\subsection}
  {\centering\normalfont\fontsize{12}{12}\bfseries} 
  {\thesubsection.}{1em}{}

\usepackage{pdflscape}
\usepackage[english]{babel}
\usepackage[latin9]{inputenc}
\usepackage[T1]{fontenc}
\usepackage{authblk}
\usepackage{ulem,url}
\usepackage{caption}
\usepackage[document]{ragged2e}
\usepackage{geometry}
\usepackage{amsmath}
\usepackage{subfig}
\usepackage{mathtools}
\usepackage{graphicx}
\usepackage[colorinlistoftodos]{todonotes}
\usepackage[colorlinks=true, allcolors=blue]{hyperref}
\usepackage{array}
\usepackage{verbatim}
\usepackage{float}
\usepackage{amsmath}
\usepackage{amssymb}
\usepackage{graphicx}
\usepackage{setspace}
\usepackage{esint}
\usepackage{graphicx} 
\newtheorem{lemma}{Lemma}
\newtheorem{corollary}{Corollary}
\newtheorem{proposition}{Proposition}
\usepackage{natbib}
 \setcitestyle{authoryear}

\title{Separating Advertising and Marketplace Functions of E-commerce Platforms: Is it Social Welfare Enhancing?}
\author[1]{Zhe Zhang}
\author[2]{Young Kwark}
\author[1]{Srinivasan Raghunathan}
\author[3]{Peng Wang}

\affil[1]{University of Texas at Dallas}
\affil[2]{University of Denver}
\affil[3]{Northwestern Polytechnical University}
\date{}

\begin{document}
\maketitle
\setstretch{1.5}

\begin{abstract}
\noindent The use of sponsored product listings in prominent positions of consumer search results has made e-commerce platforms, which traditionally serve as  marketplaces for  third-party sellers to reach consumers, a major medium for those sellers to advertise their products. 
On the other hand, regulators have expressed anti-trust concerns about an e-commerce platform's integration of marketplace and advertising functions; they argue that such integration benefits the platform and sellers at the expense of consumers and society and have proposed separating the advertising function from those platforms. We show, 
contrary to regulators' concerns, that separating the advertising function from the e-commerce platform benefits the sellers, hurts the consumers, and does not necessarily benefit the social welfare. A key driver of our findings is that an independent advertising firm, which relies solely on advertising revenue, has same or lesser economic incentive to improve targeting precision than an e-commerce platform that also serves as the advertising medium, even if both have the same ability to target consumers. This is because an improvement in targeting precision enhances the marketplace commission by softening the price competition between sellers, but hurts the  advertising revenue by softening the competition for prominent ad positions. 
\end{abstract}

\newpage
\doublespacing
\section{Introduction}
The rapid expansion in the scale and scope of e-commerce platforms (platforms, hereafter), such as Amazon and eBay, has profoundly transformed the landscape of online retailing and advertising. These platforms do more than merely connect buyers and sellers; now they also operate as dominant advertising hubs through features like sponsored product listings \citep{CNN2024GoogleSearch}. Sellers compete for ad placements to boost product visibility, with success heavily dependent on a critical factor: the precision of consumer targeting. The more accurate a platform's understanding of the consumer preferences, the better it can target ads to relevant users, thereby driving sales and maximizing profits for sellers and the platform \citep{goldfarb2011online}. On the other hand, it also raises important questions about its broader effects on pricing strategies, consumer welfare, and overall market efficiency \citep{bakos1998emerging}. 
 
\hspace{1cm}  In recent years, concerns over the concentration of power within dominant platforms have gained significant attention from regulators and policymakers. For instance, Amazon, which has substantive controls of both retail and advertising ecosystems, faces increasing scrutiny for its market dominance \citep{khan2019separation}. Consequently, there have been calls to separate its advertising business from its marketplace function  to curb monopolistic behavior, enhance competition, and protect smaller sellers \citep{Claburn2022Lawmakers}. Proponents of such separation argue that such measures could reduce conflicts of interest, foster a level playing field, and empower sellers with greater autonomy. However, these proposals often overlook critical economic repercussions \citep{dippon2022economic}, particularly the potential benefits of better consumer targeting 
and broader implications for consumer welfare and market dynamics \citep{belleflamme2015industrial}. This research note seeks to provide valuable insights to regulators in the ongoing debate about the pros and cons of their proposal to split the advertising and marketplace roles of retail platforms. 

\hspace{1cm}We derive insights with the help of a stylistic model of an online market in which two competing sellers sell their products. 
In the case where the platform integrates advertising and  marketplace functions, when a consumer visits the platform, it displays sponsored listings of sellers where the product placement is based on the sellers' bids. Conversely, in the case where the advertising business operates independently from the platform, an independent advertising firm displays sponsored listings of sellers upon a consumer's visit. Both the platform and the advertising firm have imperfect knowledge about consumer preferences for the products; however, they have the ability to improve the precision of this knowledge, thereby improving the precision of  targeting sponsored lists to consumers.

\hspace{1cm}This research note shows that the proposed separation of the advertising business from the platform may fail to achieve its intended goals.  Contrary to regulators' expectations, such a separation hurts sellers and potentially hurts the social welfare as well although it benefits consumers. The driving force for these findings is that the independent advertising firm has same or lesser incentive to improve targeting precision than an e-commerce platform that also serves as the advertising medium. An improvement in targeting precision enhances the marketplace commission by softening the price competition between sellers, but hurts the  advertising revenue by softening the competition for prominent ad positions. The social welfare can be higher under the proposed separation only when consumer preference is weak  or their likelihoods of clicking the products for eventual purchase are small. 

\hspace{1cm}This study contributes to the ongoing debate about the optimal structure of online platforms from social welfare perspective by challenging the assumption that separating the advertising business from dominant platforms will lead to better outcomes for sellers and consumers. 
Our findings offer valuable insights for regulators, platform operators, and policymakers aiming to strike a balance between market power and economic efficiency in the rapidly evolving digital marketplaces.


\section{Literature Review}
Sponsored advertising in an online marketplace has recently attracted  attention from academic community. Several studies examine the case where sellers have superior information about consumer preference compared to the marketplace \citep{long2019leveraging, choi-mela_2019mktsci}. A few papers have also examined how sellers' knowledge about consumers can change the outcomes of sponsored advertisements \citep[e.g.,][]{chen2014economic, zhang2012contextual}. For example, \citet{chen2014economic} show that an online publisher benefits from targeted advertising enabled by consumer information and that the gain increases in user heterogeneity and the number of advertisers competing for a sponsored advertising position.  \citet{zhang2012contextual} show that when an advertising intermediary enables advertisers to target consumers, it benefits by intensifying competition between advertisers, and that the number of comparison shoppers plays a critical role in determining the impact of such targeting. \citet{Kannanetal2021} study the online trading platform's profiling to decide which seller would be discovered by each buyer for a sponsored slot. 
In contrast to this stream of research where advertising is the only revenue source for the platform, we examine advertising by a platform that charges a sales commission as well. Therefore, the targeting precision affects not only sellers' competition for the ad positions but also their price competition in our context. Another key difference between our paper and the prior stream of research is that we focus on a context where the platform has consumer information and effectively shares that information with the sellers that participate in the auction for sponsored product listing, i.e., there is no information asymmetry between sellers and the marketplace when sellers place their bids. Thus, the impacts of separating the advertising business from the marketplace are driven only by the changes to the incentive for improving targeting precision.

\hspace{1cm}Early literature stream on economics of digital business models focuses on the inter-dependencies between advertising and marketplace operations in an online platform \citep{weyl2010price,brousseau2007economics}. This stream suggests integrated platforms benefit from the synergies between advertising and marketplace functionalities, such as economies of scale, enhanced network effects, and improved ad targeting due to access to marketplace data. More recent papers that focus on the economic consequence of structural separation of the advertising business from marketplaces brought about by regulations \citep{belleflamme201811} suggest a structural separation would make consumer data collection difficult,  reducing the ability of platforms to provide personalized and efficient services, whereas \citep{dippon2022economic} point out  increased market inefficiencies. \cite{khan2019separation} argues the loss of cross-subsidization could hurt consumers and sellers, because online platforms like Amazon can no longer reinvest the profit from advertising business into price reductions or promotional offers for consumers. Our paper differs from these papers by identifying the differing incentives to improve targeting precision under different structures even if there is no impact on data availability. 

\section{Model Setup}

\noindent We consider a setting in which third-party sellers use a dominant retail platform $R$  to reach their consumers. We assume two sellers $A$ and $B$, whose products are also denoted as $A$ and $B$, respectively, with no ambiguity. The platform serves not only as a sales channel or a marketplace but also as an advertising platform for the sellers. The platform charges sellers a sales commission for its marketplace services and an advertising fee for the advertising services. 

\hspace{1cm}The products are imperfect substitutes, competing in the same category. Seller $i$, $i \in \{A, B\}$, sets product price $p_i$ and pays the platform a commission of $rp_i$ on each sale, where $r \in (0,1)$ is the commission rate.  We assume a unit mass of consumers, who are heterogeneous in their preferences which are captured using Hotelling's linear city model. Thus, consumers are uniformly distributed along a unit-length line, with products $A$ and $B$ located at positions 0 and 1, respectively. A consumer at position $z \in [0,1]$ derives utility $U_A(z) = v - zt$ from product $A$, and utility $U_B(z) = v - (1-z)t$ from product $B$, where $v$ is the consumer's utility (valuation) from consuming her ideal product, and $t$ $(t>0)$ is the misfit cost, reflecting the strength or intensity of the consumer's preference. Consumers have unit demand.

\hspace{1cm}The platform offers two ad positions for sellers to advertise their products. By clicking an ad, consumers can visit the product page where they learn about the product's location and price. To focus on the advertising dynamics, we assume that consumers can learn about a product only by clicking its ad. Following the literature on internet position auctions \citep{varian2007position,edelman2007internet}, we assume each ad position has a click-through rate (CTR), reflecting the probability that a consumer clicks the ad in that position. Empirical studies indicate that CTRs differ between positions.\footnote{Differences in CTRs may stem from factors such as ad visibility \citep{jeziorski2018advertiser,ursu2018power,chen2011paid}. Factors like perceived ad quality or click cost influence consumer clicking behavior \citep{chen2014economic, RutzTrusov2011, ghose2009empirical} as well. As a result, CTR represents the expected probability of clicking an ad in a particular position.} Specifically, the top (or left) ad position typically has a higher CTR than the bottom (or right) position in vertical (or horizontal) listings \citep{lu2015position, ghose2009empirical,agarwal2011location}. We refer to the ad position with the higher CTR as the ``more-prominent position'' that has CTR $\alpha$, where $0 < \alpha \leq 1$, and the ad position with the lower CTR as the ``less-prominent position'' that has CTR $\delta$, where $0 < \delta < \alpha$.\footnote{Parameters $\alpha$ and $\delta$ are sufficient to characterize the complete clicking behavior (i.e the likelihoods of all possible clicking scenarios) of a consumer if the click events for the two positions are conditionally independent, e.g., when the click behavior is dictated  by considerations such as search budget \citep{BayeMorgan2006}.} 

\hspace{1cm}A consumers' purchase decision for a product is conditional on the consumer's click behavior and it is as follows. 
A consumer, who clicks only one (more-prominent or less-prominent) position, \textit{either} purchases the product that she clicks \textit{or} leaves the marketplace without buying any product. Specifically, if $A$ is displayed in a position and a consumer located at $z$ clicks $A$ only, then she would purchase $A$ if and only if $z\le z_A=(v-p_A)/t$, assuming outside option is zero. Conversely, if $B$ is displayed in a position and a consumer located at $z$ clicks $B$ only, then she would purchase $B$ if and only if $z > z_B=(v-p_B)/t$. If a consumer clicks both positions, then she would buy the product with a higher net utility. Thus, if a consumer located at $z$ clicks both positions, she would buy $A$ if $z\le z_{AB}$ and buy $B$ if $z>z_{AB}$, where $z_{AB}=1/2 +(p_B-p_A)/(2t)$. A consumer that does not click either would not buy any product. 

\hspace{1cm}The platform has knowledge about consumers, which provides a signal $s$ about an individual consumer's preference (i.e., location in the Hotelling line). Following  \cite{li2020informative} and \cite{KwarkChenRaghu2014reviews}, we assume the signal is perfectly informative (i.e., reveals the consumer's true location $z$) with probability $\beta$ and is uninformative otherwise. We refer to $\beta$, $0\le \beta \le 1$, as the platform's \textit{targeting precision}. Thus,  $P(s=y|z=y)=\beta$ and $P(s\neq y|z=y)=1-\beta$. The two sellers are aware of the targeting precision. The platform decides the ad positions based on the outcome of a second-price auction for the sellers. The two sellers devise a bidding strategy that automatically executes and submits bid amounts conditional on the consumer signal; that is, sellers $A$ and $B$ specify bid amounts, $b_A(s;\beta)$ and $b_B(s;\beta)$, respectively, conditional on the signal. This individual targeting is consistent with the practice, which allows advertisers or sellers to specify their bid based on demographic, browsing and  purchasing characteristics of users/consumers. The seller with the higher bid wins the auction and secures the more-prominent position, while the losing seller occupies the less-prominent position. In the case of identical bids, positions are assigned based on consumer-product match, with the product closer to the consumer's expected location placed in the more-prominent position. On a per-click basis, the auction winner pays the loser's bid amount as the advertising fee or listing fee, and the loser pays a reserve price, normalized to zero. 

\hspace{1cm}We consider the case where the platform provides both marketplace and advertising services as the benchmark because it reflects the current setup of most platforms. We refer to the benchmark as the \textit{Integrated Setup}. We also consider an alternative scenario which reflects the regulator's proposal that seeks to split the marketplace and advertising functions of the platform. We refer to the second case as the \textit{Independent Setup}, in which the platform has only the marketplace function and an independent firm/entity (advertising firm, hereafter) offers the advertising function. In the Independent Setup, the advertising firm uses position auctions, analogous to that used by the platform in the Integrated Setup. 
We assume consumer behavior, including CTRs and the bidding mechanism, remains identical across both setups to isolate the effects of the proposal to split the two functions. Moreover, we assume negligible operational and targeting precision improvement costs for the platform and advertising firm and negligible production costs for sellers.\footnote{Positive values for these costs do not change the results qualitatively.} 

\hspace{1cm}We consider a multi-stage game. In stage 1, the regulator decides whether the Integrated Setup or the Independent Setup would prevail in the market. In stage 2, the platform or the advertising firm chooses the targeting precision, depending respectively on whether the Integrated Setup or the Independent Setup is chosen in stage 1.  In stage 3, the sellers simultaneously decide their pricing and bidding strategies. In stage 4, the platform or the advertising firm receives the signal for each visiting consumer and assigns ad positions based on the auction results. Consumers click ads, make purchase decisions, and the sellers and the platform realize profits. This sequence is consistent with observations in practice.\footnote{https://advertising.amazon.com/blog/new-bidding-and-targeting-options-for-sponsored-products} 

\hspace{1cm}The information structure for the game is the following. The signal about an individual consumer's location is private knowledge of the entity (i.e., the platform or the advertising firm) that receives it, but the targeting precision is known to the sellers. Neither the platform, advertising  firm nor the sellers know an individual consumer's clicking behavior; however, they know the expected CTR for each ad position. A consumer has no knowledge about a product, without clicking on the product's ad. 
The sellers' locations are known to sellers, the platform, and the advertising firm. All other parameters, the distribution of consumer locations, and the auction rule are common knowledge. We assume a moderate valuation range, $v \in \left(t (\frac{\alpha }{(2-\alpha)  \delta}+\frac{1}{2}), 2 t\right)$ such that $z_A$, $z_B$, and $z_{AB}$ are between $(0,1)$, where $z_B<z_{AB}<z_A$. It ensures that the market is not fully covered, meaning not all consumers who click necessarily purchase.

\section{Equilibria in the Integrated and Independent Setups}
\noindent In this section, we derive the sub game perfect equilibrium using backward induction for the Integrated Setup and the Independent Setup. We first delineate consumers' purchase decisions and sellers' demand functions (Section 4.1), followed by listing and pricing decisions (Section 4.2) and the targeting precision decision (Section 4.3). 

\subsection{Consumers' Purchase Decisions and Sellers' Demand Functions}
In the last stage of the game, a consumer's purchase decision depends on her location, the products' positions in the product listing, and product prices. Suppose the consumer's location is $z$, product $A$ is displayed in the more-prominent position, and product $B$ is in the less-prominent position. From the platform's and sellers' perspectives, with probability $\alpha (1-\delta)$, this consumer would click only product $A$  and buy it only if $z\le z_A$; with probability $(1-\alpha)\delta$, she would click only product $B$  and buy it only if $z>z_B$; with probability $\alpha \delta$, she would click both product listings, and buy $A$ if $z\le z_{AB}$ or buy $B$ if $z>z_{AB}$; otherwise, she would leave the platform without any click or purchase. In the case when product $B$ is displayed in the more-prominent position, and product $A$ is in the less-prominent position, this consumer would click only product $B$ with probability $\alpha (1-\delta)$ and buy it only if $z>z_B$; click only product $A$ with probability $(1-\alpha)\delta$ and buy it only if $z\le z_A$; click both product listings with probability $\alpha \delta$, and buy $A$ if $z\le z_{AB}$ or buy $B$ if $z>z_{AB}$; or she would leave the platform without any click or purchase.

\hspace{1cm}The platform (in the Integrated Setup) or the advertising firm (in the Independent Setup) receives a signal $s$ about the consumer location $z$. We denote the probability of the consumer buying product $i$, i.e., the expected demand of product $i$, conditional on signal being $s$ and product $A$ being in the more-prominent position (denoted using  superscript $A$) as $d_i^A(s)$. Analogously, we denote the expected demand of $i$ conditional on signal being $s$ and product $B$ being in the more-prominent position (denoted using a superscript $B$) as $d_i^B(s)$. We show $d_i^A(s)$ and $d_i^B(s)$ in Table \ref{table:1D} for the following scenarios: C1 denotes $0\le s\le z_{B}$, C2 denotes $z_{B}<s\le z_{AB}$, C3 denotes $z_{AB}< s \le z_A$, and C4 denotes $z_A<s \le 1$ and show the difference in the demands between the more-prominent and less-prominent positions, i.e., $d_A^A(s)-d_A^B(s)$, for product $A$ and $d_B^B(s)-d_B^A(s)$ for product $B$ in Table \ref{table:1}. 
\begin{table}[h]
\small
\centering
\begin{tabular}{| c c c|}
\hline
$d_A^A(s)$&$d_B^A(s)$ & $s$ \\
 \hline      
 $\beta  \alpha +(1-\beta ) (\alpha  (1-\delta) z_{A}+\alpha  \delta z_{AB})$ &$(1-\beta ) ((1-\alpha ) \delta (1-z_{B})+\alpha \delta (1-z_{AB}))$  &  C1\\
$\beta \alpha+(1-\beta ) (\alpha  (1-\delta) z_{A}+\alpha  \delta z_{AB})$  &$\beta (1-\alpha )   \delta+(1-\beta ) ((1-\alpha ) \delta (1-z_{B})+\alpha  \delta (1-z_{AB}))$&   C2\\
$\beta \alpha  (1-\delta)+(1-\beta ) (\alpha  (1-\delta) z_{A}+\alpha  \delta z_{AB})$  &$\beta \delta+(1-\beta ) ((1-\alpha ) \delta (1-z_{B})+\alpha  \delta (1-z_{AB}))$&  C3\\
$(1-\beta ) (\alpha  (1-\delta) z_{A}+\alpha  \delta z_{AB})$  &  $\beta \delta+(1-\beta ) ((1-\alpha ) \delta (1-z_{B})+\alpha  \delta (1-z_{AB}))$&C4\\
\hline
\hline
$d_A^B(s)$&$d_B^B(s)$ & $s$ \\
 \hline      
$\beta  \delta+(1-\beta ) ((1-\alpha ) \delta z_A+\alpha \delta z_{AB})$& $(1-\beta ) (\alpha  (1-\delta) (1-z_{B})+\alpha  \delta (1-z_{AB}))$  & C1\\
$\beta \delta+(1-\beta ) ((1-\alpha ) \delta z_{A}+\alpha \delta z_{AB})$&$\beta \alpha  (1-\delta)+(1-\beta ) (\alpha  (1-\delta) (1-z_{B})+\alpha \delta (1-z_{AB}))$  &  C2 \\
$\beta(1-\alpha ) \delta+(1-\beta ) ((1-\alpha ) \delta z_{A}+\alpha \delta z_{AB})$& $\beta \alpha+(1-\beta ) (\alpha  (1-\delta) (1-z_{B})+\alpha \delta (1-z_{AB}))$  & C3\\
$(1-\beta ) ((1-\alpha ) \delta z_A+\alpha \delta z_{AB})$& $\beta \alpha+(1-\beta ) (\alpha  (1-\delta) (1-z_{B})+\alpha  \delta (1-z_{AB}))$  & C4\\
\hline
\end{tabular}
 \captionsetup{font=singlespacing, justification=centering}
\caption{Expected demands of products $A$ and $B$ in the case where $A$ is in the more-prominent position 
and in the case where $B$ is in the more-prominent position.}
\label{table:1D}
\end{table}

\begin{table}[H]

\small
\centering
\begin{tabular}{|c c c |}
\hline
$d_A^A(s)-d_A^B(s)$&$d_B^B(s)-d_B^A(s)$ & $s$ \\
 \hline      
$(\alpha-\delta ) ((1-\beta) z_A+\beta )$  & $(\alpha-\delta ) (1-\beta)(1-z_B)$ &  C1\\
$(\alpha-\delta) ((1-\beta) z_{A}+\beta )$ & $(\alpha-\delta)((1-\beta)(1-z_{B})+\beta)$&  C2 \\
$(\alpha-\delta) ((1-\beta) z_{A}+\beta)$& $(\alpha-\delta)((1-\beta)(1-z_{B})+\beta)$& C3\\
$(\alpha-\delta)(1-\beta)z_{A} $& $(\alpha-\delta)((1-\beta)(1-z_{B})+\beta)$&   C4\\
\hline
\end{tabular}
\captionsetup{font=singlespacing, justification=centering}
\caption{Difference in the expected demands between the more-prominent and less-prominent positions for products $A$ and $B$.}
\label{table:1}
\end{table}
Several observations from Table \ref{table:1} are worth highlighting.\\
(i) It is easy to see that, consistent with expectation, $d_A^A(s)-d_A^B(s)> 0$ and $d_B^B(s)-d_B^A(s)> 0$ for any $s$, given $0 \le \beta\le 1$ and $0<\delta<\alpha\le 1$, implying that the expected demand is higher if the product is in the more-prominent position instead of the less-prominent position.\\
(ii) It is also evident that for a seller, the difference is higher when the consumer signal is close to the seller's location than when the signal is far, i.e., $d_A^A(s)-d_A^B(s)$ is higher under scenarios C1, C2, and C3, i.e., $0\le s \le z_{A}$ than  under scenario C4, i.e., $z_{A}< s \le 1$; and $d_B^B(s)-d_B^A(s)$ is greater under scenarios C2, C3 and C4, i.e., $z_B< s \leq 1$ than  under scenario C1, i.e., $0\leq s \leq z_B$.\\
(iii) Given signal $s$, the difference in a seller's expected demands if his product is displayed in the more-prominent position and if his product is displayed in the less-prominent position depends only on his own price, since $z_A$ is a function of only $p_A$ and $z_B$ is a function of only $p_B$. Intuitively, the gain in the expected demand from the more-prominent position comes only from a consumer that clicks and considers only the product in the more-prominent position. Moreover, the loss in the expected demand when listed in the less-prominent position comes only from the consumer that clicks and considers only the product in the less-prominent position. That is, the gain or loss in demand associated with product listing, depends only on the seller's own price.\\
(iv) The difference in a seller's expected demands increases when $\alpha-\delta$ increases. This is because a consumer is more likely to click and consider only the product in the more-prominent position when $\alpha$ increases or she is less likely to click and consider only the product in the less-prominent position when $\delta$ decreases. 
 
\begin{spacing}{1}
\subsection{Sellers' Pricing and Bidding Strategy}
\end{spacing}
In both the Integrated Setup and the Independent Setup, the seller with the higher bid wins the auction, and their product is displayed in the more-prominent position. Recall that a seller's product price affects the difference in their expected demand between the two positions, thereby influencing their willingness to pay (WTP) for the more-prominent position and affecting their bidding strategy.  Sellers make their pricing and bidding decisions simultaneously in stage 3 of the game. 

\begin{lemma}
\label{biddingstrategy}
For targeting precision  $\beta$, the following prices and bids constitute an equilibrium in both the Integrated Setup and the Independent Setup. \\
(i) $p_A^*=p_B^*=\frac{\alpha \delta t+v ((1-\alpha ) \delta+\alpha  (1-\delta))+\beta  (\alpha -\delta) (t-v)}{2 (\alpha +\delta)-3 \alpha  \delta-2 \beta  (\alpha -\delta)}$;\\
(ii)
$\begin{cases}
 b_A(s)=b_{H}, b_B(s)=b_{L} & \text{if } 0 \leq s < z_B^*\\
 b_A(s)=b_B(s)=b_{H} & \text{if } z_B^* \leq s \leq z_A^*\\
 b_A(s)=b_{L}, b_B(s)=b_{H} & \text{if } z_A^* < s \leq 1\\
\end{cases}$
\\
where $b_{L}=\frac{L_1 L_2 (1-\beta ) (1-r) (\alpha -\delta)}{t (3 \alpha \delta-2 \alpha  (1-\beta )-2 (\beta +1) \delta)^2}$,
$b_{H}=\frac{L_1 L_2 (1-\beta ) (1-r) (\alpha -\delta)}{t (3 \alpha \delta-2 \alpha  (1-\beta )-2 (\beta +1) \delta)^2}+\frac{L_1 \beta  (1-r) (\alpha-\delta)}{2 \alpha  (1-\beta )-3 \alpha  \delta+2 (\beta +1) \delta}$,\\
$z_A^*=\frac{v (\alpha +\delta)-(t+v) (\beta  (\alpha -\delta)+\alpha \delta)}{t (2 \alpha  (1-\beta )-3 \alpha \delta+2 (\beta +1) \delta)}$, $z_B^*=1-\frac{v (\alpha +\delta)-(t+v) (\beta  (\alpha -\delta)+\alpha \delta)}{t (2 \alpha  (1-\beta )-3 \alpha \delta+2 (\beta +1) \delta)}$,  
$L_1=\alpha \delta (t-2 v)+v (\alpha +\delta)-\beta  (\alpha -\delta) (v-t)$ and $L_2=v (\alpha +\delta)-\alpha \delta (t+v)-\beta  (\alpha -\delta) (t+v)$.
\end{lemma}
All proofs are in the appendix unless otherwise stated.

\hspace{1cm}First, the sellers' equilibrium strategies are identical in both setups, provided the targeting precision is the same in them. This is because the auction mechanism and consumers' clicking behaviors are identical in both cases. 

\hspace{1cm}Second, we can show that $b_{H}\ge b_{L}$. As in the typical second-price auction, where the winner pays the loser's bid amount, for any $s$, bidding an amount equal to their WTP for the more-prominent position is the dominant strategy for a seller. Lemma \ref{biddingstrategy} shows that, each seller bids higher for a signal close to his own location than for a signal that is far, i.e., seller A bids $b_H$ for $0\leq s \le z_A^*$ and $b_L$ for $z_A^*< s \le 1$. Intuitively, assuming the signal is accurate,  if $0 \leq s \leq z_A^*$, the consumer may purchase product $A$; conversely, if $z_A^* < s \leq 1$, the consumer will never purchase $A$. In equilibrium, since $b_{H}\ge b_{L}$, seller $A$ wins the more-prominent position if the signal is close to its own location, i.e., $0\leq s \le z_B^*$; and symmetrically, seller $B$ wins the more-prominent position if  $z_A^*< s \leq 1$. For intermediate signals, $z_B^*< s \leq z_A^*$, both sellers bid the same amount, and the auction winner is determined by the expected misfit cost, with seller $A$ winning the more-prominent position if $z_B^*< s \leq 1/2$, and seller $B$ winning if $1/2 < s \leq z_A^*$. Sellers's product prices and their signal specific bidding strategy depends on the targeting precision. This is intuitive since more precise information about consumer location increases the likelihood of accurately predicting consumer behavior, affecting each seller's expected demand.

\begin{lemma}
\label{betaonbids}
    If targeting precision increases (i.e., $\beta$ increases), a) the low bid ($b_L$) decreases; and b) the product prices increase.
\end{lemma}

\hspace{1cm}As previously noted, sellers' bidding strategies are contingent on the signals they receive, i.e., they bid $b_H$ if the signal is close to them and bid $b_L$ if the signal is far (Lemma \ref{biddingstrategy}). A more precise signal ($\beta$ increasing) more accurately exposes the competitive advantage one seller has over the other in attracting a particular consumer. In the extreme case where the signal reveals no information about the consumer preferences ($\beta=0$), both sellers treat the consumer (in expectation) as being located at the midpoint of the line. If the prices are equal, then each seller would expect the same profit gain from being in the more-prominent position. This creates an intense bidding competition between sellers. In this case, they end up bidding the same amount for all consumers, i.e., $b_H=b_L$. As the precision increases, sellers gain more information about consumer preferences. For example, if the signal indicates a consumer close to 0, then both sellers would realize that seller $A$ has an advantage over $B$ in that $A$'s expected profit gain from being in the more-prominent position is larger than $B$'s. Recognizing the competitive disadvantage seller $B$ decreases his bid. Thus, when $\beta$ increases, low bid ($b_L$) decreases, reflecting more differentiated bidding strategies.

\hspace{1cm}Lemma \ref{betaonbids} also shows that higher targeting precision leads to softening of price competition and increased product prices. The reasons for the softening of the price competition with an increase in $\beta$ are two-fold, which can be explained by considering the two cases: (i) when a consumer clicks just one position and (ii) when a consumer clicks both positions. First, consider the case where the consumer clicks just one position. By definition, it is more likely for the consumer to click the more-prominent position than the less-prominent position. An increase in $\beta$ increases the auction winner's expected demand in this case. Since seller $A$ ($B$) would be the winner if the signal $s$ is less (greater) than 1/2, the likelihood of the consumer being located at less (greater) than $z_A$ ($z_B$) increases when $\beta$ increases if $A$ ($B$) is the winner, which increases $A$'s  ($B$'s) expected demand from the consumer if she clicks just one position. A higher expected demand induces a seller to charge a higher price. Second, in the case where the consumer clicks both positions, an increase in precision also reduces the sellers' incentives to compete on prices for the consumer. For example, if $\beta=0$, sellers would perceive that the consumer is located at the middle of the Hotelling line. A seller would expect to attract the consumer if it decreases its price slightly lower than that of its competitor, which would lead to a Bertrand-like competition. On the other hand, when the precision is positive, the consumer is more likely to be not at the middle of the line as compared to when the precision is zero. Thus, a seller would expect a reduction in price to lead to a smaller demand increase. The smaller marginal impact of price on expected demand would induce sellers to compete less intensely for the consumer when the precision increases in the case where she clicks both positions. Both these effects - the expected demand increase in the case where the consumer clicks only one position and the less intense price competition in the case where the consumer clicks both positions - contribute to an increase in seller prices when the targeting precision improves.

\begin{lemma}
\label{ratesonpricehike}
    The increase in price due to higher precision is:\\
    a) positively moderated by $\alpha$, the CTR for the more-prominent position; and \\
    b) negatively moderated by $\delta$, the CTR of the less-prominent position.
\end{lemma}

\hspace{1cm}Lemma \ref{ratesonpricehike} implies that when the price increases due to an improvement in precision, the increase is larger if consumers are more likely to click the more-prominent position (higher $\alpha$) or less likely to click the less-prominent position (lower $\delta$). The increase on the auction winner's expected demand is greater due to an increase in precision. When $\alpha$ is higher or $\delta$ is lower, consumers are more likely to just click the more-prominent position, strengthening the increase on the auction winner's expected demand. This contributes to a greater increase in seller prices when the targeting precision improves.

\subsection{Targeting  Precision Decision}

In stage 2, the platform determines the targeting precision in the Integrated Setup, whereas the advertising firm makes the targeting decision in the Independent Setup. We use subscripts $\Gamma$ and $\Delta$ to denote the Integrated Setup and Independent Setup, respectively. 

\hspace{1cm}In the Integrated Setup,  the expected profits of seller $i \in \{A, B\}$ and the platform ($R$) are computed as: 
\begin{spacing}{1}
\begin{eqnarray}
    \pi_{i,\Gamma}&=& (1-r)D_{i,\Gamma} p_{i,\Gamma}-F_{i,\Gamma},\\\
    \pi_{R,\Gamma}&=& \sum_{i \in \{A, B\}} r D_{i,\Gamma} p_{i,\Gamma}+F_{i,\Gamma}    
\end{eqnarray}
\end{spacing}
\noindent where $D_{i,\Gamma}$ is seller $i$'s expected demand, $p_{i,\Gamma}(1-r)$ is their profit per item sold, and $F_{i,\Gamma}$ is the advertising fee paid to the marketplace. 

\hspace{1cm}In the Independent Setup, the expected profits of seller $i \in \{ A, B\}$, the platform ($R$) and the advertising firm ($O$) are formulated as: 
\begin{spacing}{1}
\begin{eqnarray}
    \pi_{i,\Delta}&=&D_{i,\Delta} p_{i,\Delta} (1-r)-F_{i,\Delta},\\\
    \pi_{R,\Delta}&=&r (D_{A,\Delta} p_{A,\Delta}+D_{B,\Delta} p_{B,\Delta}), \\
     \pi_{O,\Delta}&=&F_{A,\Delta}+F_{B,\Delta} 
\end{eqnarray}
\end{spacing}
\noindent where $D_{i,\Delta}$ is seller $i$'s expected demand, $p_{i,\Delta}(1-r)$ is their profit per item sold, and $F_{i,\Delta}$ is the advertising fee paid to the advertising firm.

\begin{lemma}
\label{effectofbeta}
(i) In the Integrated Setup, the platform will choose $\beta^*_\Gamma = 0$ if $r \leq r_l$, $\beta^*_\Gamma = 1$ if  $r \geq r_h$, and
$\beta^*_\Gamma = \hat{\beta}$ if $r \in (r_l, r_h)$.\\
(ii) In the Independent Setup, the advertising firm will choose $\beta^*_\Delta = 0$\\
$r_l$, $r_h$, and $\hat{\beta}$ are provided in the proof of the Lemma.
\end{lemma}
\hspace{1cm}Lemma \ref{effectofbeta} highlights the nuanced effects of targeting precision on the platform's profitability in the Integrated Setup where the platform derives profit from two  revenue streams: sales commission and advertising fee. As shown in Lemma \ref{betaonbids}, targeting precision affects sellers' pricing and bidding strategies, which in turn influences the platform's overall profit. Greater targeting precision (higher $\beta$) amplifies the competitive advantage of one seller over another, forcing each seller to lower his bid for consumers that are expected to be far from his location. As a result, the auction winner pays a lower advertising fee (based on the losing seller's bid), and the platform's advertising fee revenue decreases as $\beta$ increases, even though the number of consumer clicks rises. We refer to this as the \textit{advertising-fees-reducing effect}. Conversely, as targeting precision improves, sellers charge higher prices, leading to increased sales commission revenue for the platform. This is due to the increased expected demand for the winning seller in the more-prominent position in the case where the consumer clicks only one position and the reduced price competition when consumers click both ads (as outlined in Lemma \ref{betaonbids}). We refer to this effect as the \textit{sales-commission-enhancement effect}.

\hspace{1cm}The commission rate plays a crucial role in the platform's decision regarding targeting precision, as it impacts the platform's two revenue streams in opposite ways.  An increase in the commission rate leads to higher sales commissions from sellers; however, it lowers sellers' willingness to pay (WTP) for the more-prominent position, thereby reducing the marketplace's advertising revenue. Consequently, at low (high) commission rates, the \textit{advertising-fees-reducing effect} (\textit{sales-commission-enhancement effect}) dominates. For moderate commission rates, the overall impact of targeting precision on the platform's profit depends on the balance between these two opposing effects. 


\hspace{1cm}Lemma \ref{effectofbeta} shows that the platform may have little incentive to improve targeting precision when it cannot charge a high commission rate in the Integrated Setup. Note that this disincentive exists even if improving targeting precision incurs no additional cost for the platform. Moreover, a similar disincentive exists for the advertising firm in the Independent Setup because the advertising firm derives revenue solely from advertising fees and is affected by only the negative \textit{advertising-fees-reducing effect}.  

\begin{corollary}
\label{incentive}
The targeting precision in the Integrated Setup is not less than that in the Independent Setup.
\end{corollary}

\hspace{1cm}Corollary \ref{incentive} underscores a subtle, but a significant, consequence of spinning off the advertising business from a dominant retail platform, such as Amazon or eBay; it diminishes the incentive to improve the targeting precision. Even if consumer behavior and the auction mechanism remain unchanged in the two setups, the disparity in targeting precision will lead to differences in sellers' pricing and bidding decisions.

\section{Impact of Separating Advertising from the Platform}
We analyze the effects of separating advertising functions from the marketplace by comparing the seller payoffs, consumer surplus, and the social welfare in the Integrated Setup and Independent Setup. It is easy to verify that the platform can never be better off in the Independent Setup compared to the Integrated Setup which is intuitive. 

\begin{proposition}
\label{twocasecompare}
Compared to the Integrated Setup, in the Independent Setup, \\(i) sellers' profits are not higher; \\
(ii) consumer surplus is not lower; and\\
(iii) social welfare is lower  if $r > r_l$, $t>(3-\sqrt{6})v$ and $\delta>\frac{4 t}{3 v-t}$.
\end{proposition}

\hspace{1cm}Proposition \ref{twocasecompare} is the central result of this paper and provides answers to our research questions. It underscores the potential unintended consequences of splitting the advertising and the marketplace roles of a platform. Though this action could help consumers at large, it hurts sellers. This action hurts the society if the commission rate is not too low, unit misfit cost (which measures product differentiation or inverse of price competition in the market) is high, and CTRs of ad positions are also high. The following corollary reveals that the key driver of the unintended consequences of splitting the advertising function and the marketplace function of a platform is its impact on the incentive to improve the targeting precision.

\begin{corollary}
\label{identity}
    Sellers' profits, consumer surplus, and social welfare in the Independent Setup are identical respectively to those in the Integrated Setup if the targeting precision is identical, not necessarily zero,  in the two cases. 
\end{corollary}

\hspace{1cm}If splitting the advertising business (i.e., shift from the Integrated Setup to the Independent Setup) does not alter the targeting precision, then there is no impact from it. Recall from Lemma 3 that targeting precision remains the same when advertising function is split from the platform when the commission rate is low (i.e., $\beta^*_\Gamma=\beta^*_\Delta=0$ when $r \le  r_l$). In this case, sellers' profits, consumer surplus, and social welfare remain the same after the split. On the other hand, if splitting the advertising business means that the advertising firm would alter (reduce, in our context) the targeting precision, then the sellers are worse off and the society can also be worse off. Recall again from Lemma 3 that targeting precision reduces when advertising function is split from the platform when the commission rate is not too low (i.e., $\beta^*_\Gamma>0$ and $\beta^*_\Delta=0$ when $r > r_l$). Essentially, the advertising firm's and the platform's differential incentives to improve targeting precision contribute to the potential unintended consequences stated in Proposition \ref{twocasecompare}. 

\hspace{1cm}A reduction in the targeting precision intensifies price competition between the sellers, decreasing their profits from sales. Moreover, it exacerbates the competition for ads positions, driving up the listing fees sellers pay for advertising and thereby increasing their costs. As a result, sellers would earn lower profit if the advertising business is separated from the platform.  
On the other hand, the same effects of a decrease in targeting precision are beneficial to consumers. Clearly, the lower product prices resulting from intensified price competition benefits consumers. Although lower precision means the marketplace is less capable of identifying consumer preferences and thereby displaying products more suited to consumers (i.e., closer to their locations) in the more-prominent positions, the surplus gain due to the price drop outweighs the increase in consumers' misfit costs. 
Thus, if the advertising business is separated from the platform, consumer surplus increases, as the lower precision works in favor of consumer welfare.
  
\hspace{1cm}When the commission rate is not too low ($r > r_l$), the trade-off of reduced targeting precision due to separating the advertising business from the platform becomes clear: consumers enjoy higher surplus, but sellers' profits reduce. From a social planner's perspective, it is essential to assess the overall impacts of the separation of the advertising business from the perspective of social welfare that includes the marketplace's and the advertising firm's profits, sellers' profits, and the consumer surplus.

\hspace{1cm}Essentially, the social welfare represents the expected utility that consumers derive from the marketplace because prices and advertising fees are simply transfers of money between different players within the society. When targeting precision decreases, it is likely that the product displayed in the more-prominent position does not match with the consumer's preference (far from the consumer's location), increasing the misfit cost and reduces the consumer's expected utility. This is the direct negative effect. However, lower precision also allows the sellers to reduce prices, which enables more consumers to buy a product, thus increasing consumer utility. 
If the CTR for the more-prominent position is high and the unit misfit cost is high  (i.e., large $\alpha$ and $t$) , then the direct negative effect is strong since consumers are likely to click on the ill-matched product displayed at the more-prominent position and incur higher misfit cost. Moreover, the CTR for the less-prominent position negatively affects the price drop from a decrease in precision (i.e., larger $\delta$ means smaller price alteration due to changes in targeting precision based on Lemma \
\ref{ratesonpricehike}). If the CTR for the less-prominent position is high (i.e., large $\delta$), then the indirect positive effect about the price drop is weakened. As a result, when $t$, $\alpha$ and $\delta$ are large, i.e., $t>(3-\sqrt{6})v$ and $\alpha>\delta>\frac{4 t}{3 v-t}$ , the direct negative effect dominates, leading to a drop in social welfare as precision decreases. Thus, given Proposition \ref{twocasecompare} part (iii), social welfare decreases when the advertising business is separated from the platform if the commission rate is not too low, and CTRs of the two ad positions are high.

\hspace{1cm}On the other hand,  can splitting the advertising business enhance social welfare when the ad positions' CTR or commission rate is low? Furthermore, how does the intensity of consumer preference ($t$) affect the impact of regulator's proposal to split the advertising business on social welfare? We provide insights into these questions primarily through numerical observations as theoretical analysis proved to be challenging.

It is straightforward to verify that the social welfare remains the same in the Integrated Setup and the Independent Setup when $r\le r_l$ because the optimal targeting precision is zero in both setups. 
When the commission rate is not low ($r> r_l$) and 
the CTR of both the ad positions are low, i.e., $\delta<\alpha<\frac{4 t}{3 v-t}$, we observe the following. 

\noindent\textbf{Observation 1}
\textit{The social welfare can be higher in the Independent Setup compared to the Integrated Setup if unit of misfit cost is small ($t$) or CTRs of ad positions ($\alpha$ are $\delta$) are small.}


\begin{figure}[h]
  \centering
    \includegraphics[width=6in]{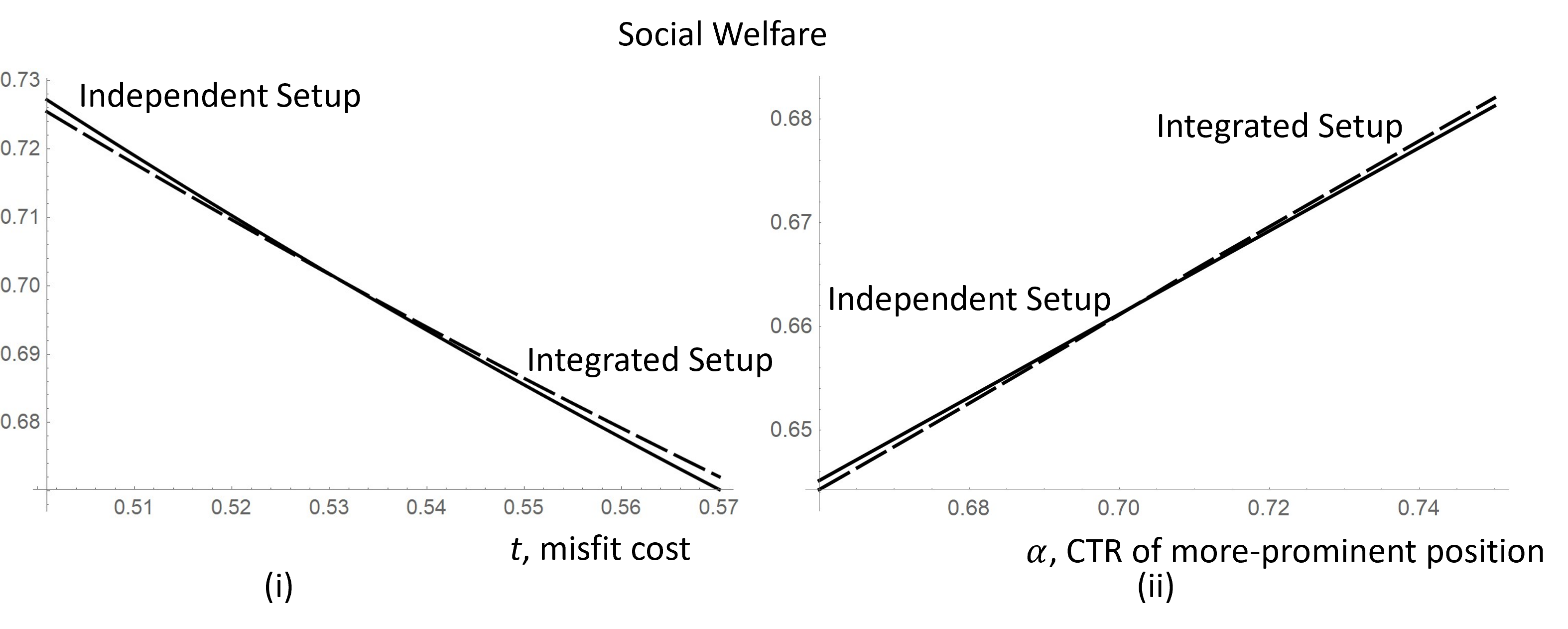}
    \caption{Social welfare given the optimal precision in the Integrated Setup (dashed line) and the Independent Setup (solid line), i) with increasing $t$ where $v=1 ,r=0.25, \alpha=0.7, \delta=0.65$; ii) with increasing $\alpha$ where  $v=1 ,r=0.35, \delta=0.55, t=0.54$}
      \label{swbeta2}
\end{figure}

\hspace{1cm}Figure \ref{swbeta2} illustrates the above observation. In both plots, the solid line represents social welfare in the Independent Setup and the dashed line represents social welfare in the Integrated Setup. In the figure, social welfare is higher in the Independent Setup than in the Integrated Setup when (i) the consumer preference is weak (the misfit cost $t$ is small); and when (ii) CTR of the more-prominent positions ($\alpha$) is low. In both plots, the CTR of the less-prominent positions ($\delta$) is not high.  The intuitions of this promising observation that supports the regulators' contention  are twofold. First, when $t$ and $\alpha$ are low, consumers are less likely to click the more-prominent position and incur a small misfit cost, thus, the direct negative effect of reduced precision on social welfare weakens. Second, when $\delta$ is not high, it is very likely that consumers only click on a product that is far away from their locations when the signals about them are incorrect. A price drop due to reduced precision (with a transition from Integrated Setup to Independent Setup) makes them more likely to complete a purchase, strengthening the indirect positive effect of reduced precision on social welfare. Therefore, the social welfare increases under those conditions. 

\section{Conclusion}
The findings of this study underscore the critical role of integrated advertising and marketplace operations in enhancing efficiency, aligning seller and platform incentives, and maximizing social welfare. While calls to separate advertising businesses from dominant e-commerce platforms aim to curb monopolistic practices and foster competition, this research note demonstrates that such separation may lead to unintended consequences, including reduced precision in consumer targeting, misaligned incentives, and inefficiencies in marketplace dynamics. The study highlights that integrated platforms, despite potential trade-offs, are better equipped to leverage consumer knowledge for personalized experiences, efficient ad placements, and competitive seller opportunities. Policymakers should carefully consider these dynamics to avoid regulatory measures that may inadvertently harm sellers, consumers, and overall market efficiency. 

\begin{spacing}{1}
\setlength{\bibsep}{5pt} 
\bibliographystyle{ormsv080}
\bibliography{litbib}
\end{spacing}

\section*{Appendix}

\begin{spacing}{1}
\textbf{Proof of Lemma \ref{biddingstrategy}}


In stage 1, both sellers have the following common belief about the bidding strategies: For a consumer with signal $s$, seller $A$ bids more than seller $B$, i.e., $b_A(s)\ge b_B(s)$, if  $0\leq s \le z_B$; seller $A$ bids the same amount as seller $B$, i.e., $b_A(s)=b_B(s)$, if $z_B\leq s\leq z_A$; and seller $A$ bids lower than seller $B$, i.e., $b_A(s)\le b_B(s)$, if $z_A<s\leq 1$. We show subsequently that this belief and equilibrium strategies are consistent with each other.

Based on the auction rule, if the two bids are identical, seller $A$ wins the auction if $z_B< s \leq 1/2$ due to higher product-consumer fit. Similarly, $B$ wins the auction if $1/2 < s\leq z_A$. Thus, seller $A$ wins the auction 
and pays $b_B(s)$ if $0\leq s \leq 1/2$ and the consumer clicks; seller $B$ wins the auction
and pays $b_A(s)$ if $1/2 < s\leq 1$ and the consumer clicks. 

A seller's bid amount is upper bounded by his willingness-to-pay for the more-prominent position,  which is equal to the surplus gain he would get by occupying more-prominent position relative to the less-prominent position. For each seller, a bid amount equal to his willingness-to-pay weakly dominates any other bid. Therefore, by eliminating dominated bids, we find that each seller bidding an amount equal to his willingness-to-pay for the more-prominent position constitutes an equilibrium. That is, $b_A(s)=(d_A^A-d_A^B)p_A(1-r)$ and $b_B(s)=(d_B^B-d_B^A)p_B(1-r)$.

From Table \ref{table:1}, we obtain the sellers' bid amounts, which depend on prices, as follows: 

When $0\leq s \le z_{B}$: $b_A(s)=b_{A,H}=(\alpha-\delta)  \left(\beta +\frac{(1-\beta) (v-p_A)}{t}\right) p_A(1-r)$ and $b_B(s)=b_{B,L}=(\alpha-\delta) \frac{  (1-\beta) (v-p_B)}{t} p_B(1-r)$.\\
When $ z_{B} < s \leq z_{A}$:
$b_A(s)=b_{A,H}$ and $b_B(s)=b_{B,H}=(\alpha-\delta) \left(\beta +\frac{(1-\beta) (v-p_B)}{t}\right) p_B(1-r)$.\\
When $ z_{A} < s \leq 1$: $b_A(s)=b_{A,L}=(\alpha-\delta) \frac{  (1-\beta) (v-p_A)}{t} p_A(1-r)$ and  $b_B(s)=b_{B,H}$.

Suppose $p_A\le p_B$, then $z_{AB}\le 1/2$. We write the sellers' expected profits as follows:

$E(\pi_A)= \left (\int_{0}^{1/2} d_A^A ds + \int_{1/2}^{1} d_A^B ds \right )p_A(1-r)-\alpha \int_{0}^{1/2} b_B(s) ds$\\
$=\left(\substack{\int_{0}^{z_{AB}} \beta\alpha+(1-\beta)(\alpha (1-\delta)z_A+\alpha \delta z_{AB}) ds+\int_{z_{AB}}^{1/2} \beta \alpha (1-\delta)+(1-\beta)(\alpha (1-\delta)z_A+\alpha \delta z_{AB}) ds+\\ \int_{1/2}^{z_{A}} \beta(1-\alpha)\delta+(1-\beta)((1-\alpha)\delta z_{A}+\alpha \delta z_{AB} ) ds+\int_{z_{A}}^{1} (1-\beta)((1-\alpha)\delta z_{A}+\alpha \delta z_{AB} ) ds} \right) p_A(1-r)$\\ $-\alpha \left(\int_{0}^{z_{B}} b_{B,L} ds+\int_{z_{B}}^{1/2} b_{B,H} ds \right)$\\ 
$E(\pi_B)= \left(\int_{0}^{1/2} d_B^A ds + \int_{1/2}^{1} d_B^B ds\right)p_B(1-r)-\alpha\int_{1/2}^{1} b_A(s) ds$\\
$=\left( \substack{\int_{0}^{z_{B}} (1-\beta)((1-\alpha)\delta(1-z_B)+\alpha \delta(1-z_{AB})) ds +\int_{z_{B}}^{z_{AB}} \beta(1-\alpha)\delta+(1-\beta)((1-\alpha)\delta(1-z_B)+\alpha\delta (1-z_{AB})) ds\\+\int_{z_{AB}}^{1/2} \beta\delta+(1-\beta)((1-\alpha)\delta(1-z_B)+\alpha\delta (1-z_{AB})) ds+\int_{1/2}^{1} \beta \alpha+(1-\beta)(\alpha (1-\delta)(1-z_B)+\alpha\delta(1-z_{AB}))ds } \right)p_B(1-r)$ \\$-\alpha\left(\int_{1/2}^{z_A} b_{A,H} ds+\int_{z_{A}}^{1} b_{A,L} ds \right)$

After substituting $z_A$, $z_B$, and $z_{AB}$, we take the derivative of $E(\pi_A)$ with respect to $p_A$ and the derivative of $E(\pi_B)$ with respect to $p_B$ and get:\\
$\frac{d E(\pi_i)}{d p_i}=\frac{(1-r) (L_1-p_i (2 \alpha  (1-\beta )+2 \delta (1-\alpha +\beta))+\alpha  \delta -p_{-i})}{2 t}$, where $i \in \{A, B\}$, $L_1=\alpha \delta (t-2 v)+v (\alpha +\delta)-\beta  (\alpha -\delta) (v-t)>0$. Given $0<\delta<\alpha<1,0 \le \beta\le 1, t>0, t \left(\frac{\alpha }{2 \delta-\alpha \delta}+\frac{1}{2}\right)<v$, we have $L_1>0$.

Solving the first order conditions simultaneously, we get the optimal prices $p_A^*=p_B^*=\frac{\alpha \delta t+v ((1-\alpha ) \delta+\alpha  (1-\delta))+\beta  (\alpha -\delta) (t-v)}{2 (\alpha +\delta)-3 \alpha  \delta-2 \beta  (\alpha -\delta)}$.

We check the second order condition $\frac{d^2 E(\pi_A)}{d p_A^2}=\frac{d^2 E(\pi_B)}{d p_B^2}=-\frac{(1-r) (\alpha  (1-\beta )+\delta (1-\alpha +\beta))}{t}<0$, given $0 \le \beta \le 1, 0<r<1, 0<\delta<\alpha<1, t>0$.

With $p_A^*$ and $p_B^*$, $b_{A,H}({p_A^*})=b_{B,H}({p_B^*})=b_{H}=\frac{L_1 L_2 (1-\beta ) (1-r) (\alpha -\delta)}{t (3 \alpha \delta-2 \alpha  (1-\beta )-2 (\beta +1) \delta)^2}+\frac{L_1 \beta  (1-r) (\alpha-\delta)}{2 \alpha  (1-\beta )-3 \alpha  \delta+2 (\beta +1) \delta}$ \\
$b_{A,L}({p_A^*})=b_{B,L}({p_B^*})=b_{L}=\frac{L_1 L_2 (1-\beta ) (1-r) (\alpha -\delta)}{t (3 \alpha \delta-2 \alpha  (1-\beta )-2 (\beta +1) \delta)^2}$, where $L_2=v (\alpha +\delta)-\alpha \delta (t+v)-\beta  (\alpha -\delta) (t+v)$.

Given $0<\delta<\alpha<1,0 \le \beta\le 1, t>0, t \left(\frac{\alpha }{2 \delta-\alpha \delta}+\frac{1}{2}\right)<v$, we have $L_2>0$. We have $\frac{L_1 L_2 (1-\beta ) (1-r) (\alpha -\delta)}{t (3 \alpha \delta-2 \alpha  (1-\beta )-2 (\beta +1) \delta)^2}\ge 0$, given $L_1>0, L_2>0$, $0<\delta<\alpha<1,0 \le \beta\le 1, t>0$ and $0<r<1$. Thus, $b_L\ge 0$.

We get $\frac{L_1 \beta  (1-r) (\alpha-\delta)}{2 \alpha  (1-\beta )-3 \alpha  \delta+2 (\beta +1) \delta}>0$, given $t \left(\frac{\alpha }{2 \delta-\alpha \delta}+\frac{1}{2}\right)<v$, $0<\delta<\alpha<1$, $L_1>0, 0<r<1$ and $0 \le \beta \le 1$, thus, $b_{H}>0$ and $b_{H}-b_{L}\ge0$. Therefore, $b_A(s)\ge b_B(s)$ if  $0 \leq s \le z_B$, $b_A(s)=b_B(s)$ if  $z_B < s\leq z_A$, and $b_A(s)\le b_B(s)$ if  $z_A<s \leq 1$.

We also get $z_{AB}^*=z_{AB}({p_A^*,p_B^*})=1/2$, $z_A^*=z_A(p_A^{*})=\frac{v (\alpha +\delta)-(t+v) (\beta  (\alpha -\delta)+\alpha \delta)}{t (2 \alpha  (1-\beta )-3 \alpha \delta+2 (\beta +1) \delta)}$ and $z_B^*=z_B(p_B^{*})=1-\frac{v (\alpha +\delta)-(t+v) (\beta  (\alpha -\delta)+\alpha \delta)}{t (2 \alpha  (1-\beta )-3 \alpha \delta+2 (\beta +1) \delta)}$.

Given $t \left(\frac{\alpha }{2 \delta-\alpha \delta}+\frac{1}{2}\right)<v<2 t$, $0 \le \beta \le 1$, $t>0$ and $0<\delta<\alpha<1$, $0<z_B^*<1/2<z_A^*<1$. $\blacksquare$

\textbf{Proof of Lemma \ref{betaonbids}}

(a) From Lemma \ref{biddingstrategy}, $b_{L}=\frac{L_1 L_2 (1-\beta ) (1-r) (\alpha -\delta)}{t (3 \alpha \delta-2 \alpha  (1-\beta )-2 (\beta +1) \delta)^2}$, where $L_1=\alpha \delta (t-2 v)+v (\alpha +\delta)-\beta  (\alpha -\delta) (v-t)$ and $L_2=v (\alpha +\delta)-\alpha \delta (t+v)-\beta  (\alpha -\delta) (t+v)$. 

$\frac{\partial L_1}{\partial \beta}= -(\alpha -\delta) (v-t)<0$ and $\frac{\partial L_2}{\partial \beta}= -(\alpha -\delta) (t+v)<0$, given $0<t \left(\frac{\alpha }{2 \delta-\alpha \delta}+\frac{1}{2}\right)<v<2 t$ and $0<\delta<\alpha<1$. Also, we showed that  $L_1>0$ and $L_2>0$ in Proof of Lemma \ref{biddingstrategy}. Thus, it is easy to see that the numerator of $b_{L}$ decreases when $\beta$ increases, since all three parts, i.e., $L_1$, $L_2$ and  $(1-\beta ) (1-r) (\alpha -\delta)$, are positive and decrease when $\beta$ increases. 

We can rewrite the term in the denominator of $b_L$ as  $t((4-3 \alpha ) \delta+2 (1-\beta ) (\alpha -\delta))^2$, where $-(3 \alpha \delta-2 \alpha  (1-\beta )-2 (\beta +1) \delta)=(4-3 \alpha ) \delta-2 (1-\beta ) (\alpha -\delta)$ is positive and increases when $\beta$ increases, given   $t \left(\frac{\alpha }{2 \delta-\alpha \delta}+\frac{1}{2}\right)<v<2 t$ , $0\le \beta \le 1$ and $0<\delta<\alpha<1$. Thus, the denominator of $b_L$ increases when  $\beta$ increases, given  $0<t$, $0\le \beta \le 1$ and $0<\delta<\alpha<1$. 
Hence, $b_{L}$ decreases when $\beta$ increases.

(b) 
$\frac{\partial p^*_A}{\partial\beta}=\frac{(\alpha -\delta) (\alpha  (2-\delta) t+\delta (2 t-\alpha  v))}{(2 \alpha  (\beta -1)+3 \alpha  \delta-2 (\beta +1) \delta)^2}$. Given $0<v<2 t$, $0 \le \beta \le 1$, $t>0$ and $0<\delta<\alpha<1$,  the numerator of $\frac{\partial p^*_A}{\partial\beta}$ is positive, whereas it is easy to see its denominator is also positive. Thus, $\frac{\partial p^*_A}{\partial\beta}>0$. $\blacksquare$

\textbf{Proof of Lemma \ref{ratesonpricehike}}\\
(a) 
$\frac{\partial^2 p^*_A}{\partial \beta \partial \alpha}=\frac{\delta \left(\delta^2 (3 \alpha  t+2 (\beta -5) t+v (3 \alpha +2 \beta +2))-2 \delta (t ((\alpha +4) \beta +3 \alpha -4)+\alpha  (\beta +3) v)+8 \alpha  (\beta +1) t\right)}{(2 (\beta +1) \delta+2 \alpha  (1-\beta )-3 \alpha  \delta)^3}$. \\
We take the derivative of the second term of the numerator of $\frac{\partial^2 p^*_A}{\partial \beta \partial \alpha}$,  \\$\left(\delta^2 (3 \alpha  t+2 (\beta -5) t+v (3 \alpha +2 \beta +2))-2 \delta (t ((\alpha +4) \beta +3 \alpha -4)+\alpha  (\beta +3) v)+8 \alpha  (\beta +1) t\right)$, w.r.t. $v$, and get $-2 \beta  \delta (\alpha -\delta)-\delta (\alpha +2 (\alpha -\delta)+3 \alpha  (1-\delta))<0$, given $0 \le \beta \le 1$ and $0<\delta<\alpha<1$. This means the second term of the numerator decreases with $v$. Then, we evaluate this term at the maximum of $v$, $v=2t$, and get $2 \beta  (4-3 \delta) t (\alpha -\delta)+(4-3 \delta) t ((2-\alpha ) \delta+2 \alpha  (1-\delta))>0$, given $t>0$, $0 \le \beta \le 1$ and $0<\delta<\alpha<1$. Since this term it is decreasing with $v$ and is positive at the maximum of $v$, this term is positive given $t<2v$, $0 \le \beta \le 1$ and $0<\delta<\alpha<1$. Thus, the numerator of $\frac{\partial^2 p^*_A}{\partial \beta \partial \alpha}$ is positive.

We proved that $2 \alpha  (1-\beta )-3 \alpha  \delta+2 (\beta +1) \delta>0$ given $0\le \beta \le 1$ and $0<\delta<\alpha<1$ in Proof of Lemma 1. Thus, the denominator of $\frac{\partial^2 p^*_A}{\partial \beta \partial \alpha}$ is positive.
Therefore, $\frac{\partial^2 p^*_A}{\partial \beta \partial \alpha}>0$.

(a) 
$\frac{\partial^2p^*_A}{\partial \beta \partial \delta}=-\frac{\alpha  \left(\delta t (8-\alpha  (-3 \alpha -2 \beta +6)-8 \beta )-v (\alpha  \delta-2 \alpha  (1-\beta ) (\alpha -\delta)+3 (1-\alpha ) \alpha  \delta)-2 \alpha ^2 (\beta +5) t+8 \alpha  (\beta +1) t\right)}{(2 \alpha  (1-\beta )-3 \alpha  d+2 (\beta +1) \delta)^3}$\\
We take the derivative of the second term of the numerator of $\frac{\partial^2 p^*_A}{\partial \beta \partial \delta}$ w.r.t. $v$, and get $-(\alpha  \delta-2 \alpha  (1-\beta ) (\alpha -\delta)+3 (1-\alpha ) \alpha  \delta)<0$, given $0 \le \beta \le 1$ and $0<\delta<\alpha<1$. This means the second term of the numerator decreases with $v$. Then, we evaluate this term at the maximum of $v$, $v=2t$, and get $2 (4-3 \alpha ) \beta  t (\alpha -\delta)+(4-3 \alpha ) t ((2-\alpha ) \delta+2 \alpha  (1-\delta))>0$, given $t>0$,  $0 \le \beta \le 1$ and $0<\delta<\alpha<1$. Since this term is positive at the maximum of $v$ and it is decreasing with $v$, this term is positive given $t<2v$, $0 \le \beta \le 1$ and $0<\delta<\alpha<1$. Thus, the numerator of  $\frac{\partial^2 p^*_A}{\partial \beta \partial \delta}$ is positive.

We proved that $2 \alpha  (1-\beta )-3 \alpha  \delta+2 (\beta +1) \delta>0$ given $0\le \beta \le 1$ and $0<\delta<\alpha<1$. Thus, the denominator of  $\frac{\partial^2 p^*_A}{\partial \beta \partial \delta}$ is positive.
Thence,  $\frac{\partial^2 p^*_A}{\partial \beta \partial \delta}<0$, because its numerator and denominator are positive, and the sign before the fraction is negative. $\blacksquare$

\textbf{Proof of Lemma \ref{effectofbeta}}

(i) Given sellers' bids $b_H$ and $b_L$ and prices $p_A^{*}$ and  $p_B^{*}$ from Lemma \ref{biddingstrategy}, in the Integrated Setup, the marketplace's sale commission is $E(\pi_{C,\Gamma})= r p_A^*(\int_{0}^{1/2} d_A^A ds+ \int_{1/2}^{1} d_A^B ds)+r p_B^* ( \int_{0}^{1/2} d_B^A ds+ \int_{1/2}^{1} d_B^B ds)=\frac{r L_1^2(\alpha  (1-\beta )+\delta (1-\alpha +\beta))}{t (3 \alpha  \delta-2 \alpha  (1-\beta )-2 (\beta +1) \delta)^2}$, and the marketplace's product listing fee is \\$E(\pi_{M,\Gamma})=\alpha\left(\int_{0}^{z_B^*} b_L ds+\int_{z_B^*}^{1/2} b_H ds+\int_{1/2}^{z_A^*} b_H ds+\int_{z_A^*}^{1} b_L ds\right)=\frac{(1-r) L_1 L_3\alpha (\alpha -\delta)}{t (3 \alpha  \delta-2 \alpha  (1-\beta )-2 (\beta +1) \delta)^2}$\\where $L_3=\alpha  \left(v(1-\beta ^2)-(3-\beta ) \beta  t\right)-\delta \left(t \left(\alpha(1-2 \beta) +\beta (1+\beta) \right)-(\beta +1) v (1-\alpha +\beta)\right)$.

Given $0<\delta<\alpha<1,0 \le \beta\le 1, t>0, t \left(\frac{\alpha }{2 \delta-\alpha \delta}+\frac{1}{2}\right)<v$, we have $L_3>0$.

Hence, in the Integrated Setup, the marketplace's expected profit is  \\$E(\pi_{R,\Gamma})=E(\pi_{C,\Gamma})+E(\pi_{M,\Gamma})=\frac{L_1 \left((1-r)L_3 \alpha(\alpha-\delta)+r L_1(\alpha(1-\beta)+\delta (1-\alpha+\beta))\right)}{t (3 \alpha  \delta-2 \alpha  (1-\beta )-2 (\beta +1) \delta)^2}$

This is evident that, since $L_3>0$ and we have proved $L_1>0$ (proof of Lemma 1), and also given $0<\delta<\alpha<1,0 \le \beta\le 1, 0<r<1, t>0$, we have $E(\pi_{M,\Gamma})>0$ and $E(\pi_{C,\Gamma})>0$, and thus, $E(\pi_{R,\Gamma})>0$. 

$ \frac{\partial E(\pi_{R,\Gamma})}{\partial \beta}=\frac{\alpha  N_1 (1-r) (\alpha -\delta)+L_1 N_2 r}{t (3 \alpha \delta-2 \alpha  (1-\beta )-2 (\beta +1) \delta)^3}$,
where  $L_1'=\frac{\partial L_1}{\partial \beta }$, $N_2'=\frac{\partial N_2}{\partial \beta }$, $N_1'=\frac{\partial N_1}{\partial \beta }$, $N_2=2L_1 (\alpha  (1-\beta )+\delta (1-\alpha +\beta )) (3 \alpha  \delta-2 \alpha  (1-\beta )-2 (\beta +1) \delta)+L_1' (\alpha -\delta) (\delta (\alpha -2 (\beta +1))-2 \alpha  (1-\beta ))$ and 
$N_1=L_3 L_1' (3 \alpha  \delta-2 \alpha  (1-\beta )-2 (\beta +1) \delta)+L_1 L_3' (3 \alpha  \delta-2 \alpha  (1-\beta )-2 (\beta +1) \delta)-4 L_1  L_3  (\alpha -\delta)$.\\
Note that $N_1$, $L_1$ and $N_2$ are not functions of $r$. 

Solving $\frac{\partial E(\pi_{R,\Gamma})}{\partial \beta}=0$, we get $r=\frac{\alpha  N_1 (\alpha -\delta)}{\alpha  N_1 (\alpha -d)-N_1 N_2}$. 

$\frac{\partial ^2E(\pi_{R,\Gamma})}{\partial \beta^2}=\frac{(3 \alpha \delta-2 \alpha  (1-\beta )-2 (\beta +1) \delta) \left(N_2 r L_1'+L_1 r N_2'+\alpha  (1-r) (\alpha -\delta) N_1'\right)-6 (\alpha -\delta) (L_1 N_2 r+\alpha  N_1 (1-r) (\alpha -\delta))}{t (3 \alpha  \delta-2 \alpha  (1-\beta)-2 (\beta +1) \delta)^4}$.

By equating $\frac{\partial^2 E(\pi_{R,\Gamma})}{\partial \beta^2}=0$, we get $r=r_{h1}$\\$=\frac{\alpha  \left(\delta^2 (2 (\beta +1) t-\alpha  t (11-3 \alpha -\beta)+\alpha  v (3 \alpha +\beta +1))-\alpha  \delta (\alpha  (\beta +5) v-t (8-(\alpha +8) \beta -5 \alpha))+6 \alpha ^2 (\beta +1) t\right)}{\delta^2 (t (2 (\beta +1)-\alpha  (\alpha  (8-3 \alpha -\beta)+\beta +1))+\alpha  v (1-\alpha  (2-3 \alpha -\beta)+\beta))+\alpha  \delta (t (\alpha  (5-\alpha \beta-5 \alpha -5 \beta)+4)+\alpha  v (1-\alpha  (\beta +5)-\beta))+2 \alpha ^2 t (3 \alpha  (\beta +1)-\beta +1)}$. We verify that when $r<r_{h1}$, $\frac{\partial^2 E(\pi_{R,\Gamma})}{\partial \beta^2}<0$ for $0\le\beta\le 1$. 

Moreover, by equating $\frac{\partial E(\pi_{R,\Gamma})}{\partial \beta}|_{\beta=0}=0$, we get $r=r_l=\frac{\alpha  N_1 (\alpha -\delta)}{\alpha  N_1 (\alpha -d)-N_1 N_2}|_{\beta=0}$. We verify that  $\frac{\partial E(\pi_{R,\Gamma})}{\partial \beta}|_{\beta=0}>0$  if $r>r_l$. By equating $\frac{\partial E(\pi_{R,\Gamma})}{\partial \beta}|_{\beta=1}=0$, we get $r=r_{h2}=\frac{\alpha  N_1 (\alpha -\delta)}{\alpha  N_1 (\alpha -d)-N_1 N_2}|_{\beta=1}$. 

We verify that $\frac{\partial E(\pi_{R,\Gamma})}{\partial \beta}|_{\beta=1}<0$, if $r<r_{h2}$ and  $r_l<r_{h1}$. We define $r_h=\text{min}(r_{h1},r_{h2})$. 

If $r\le r_l$, then $\frac{\partial^2 E(\pi_{R,\Gamma})}{\partial \beta^2}<0$ and $\frac{\partial E(\pi_{R,\Gamma})}{\partial \beta}|_{\beta=0}<0$, means that $E(\pi_{R,\Gamma})$ decreases with $\beta$, given $0\le \beta \le 1$.

If $r\ge r_h$, then 
$\frac{\partial E(\pi_{R,\Gamma})}{\partial \beta}|_{\beta=0}>0$, and either 1) $\frac{\partial^2 E(\pi_{R,\Gamma})}{\partial \beta^2}>0$  or 2) $\frac{\partial^2 E(\pi_{R,\Gamma})}{\partial \beta^2}<0$ and $\frac{\partial E(\pi_{R,\Gamma})}{\partial \beta}|_{\beta=1}>0$. In either case, $E(\pi_{R,\Gamma})$ increases with $\beta$, given $0\le \beta \le 1$.

If $r_l<r<r_h$, then $\frac{\partial^2 E(\pi_{R,\Gamma})}{\partial\beta^2}<0$, $\frac{\partial E(\pi_{R,\Gamma})}{\partial \beta}|_{\beta=0}>0$ and $\frac{\partial E(\pi_{R,\Gamma})}{\partial \beta}|_{\beta=1}<0$.
Thus, there exists an optimal $\beta=\hat{\beta}$, that $0<\hat{\beta}<1$, which satisfies $\frac{\partial E(\pi_{R,\Gamma})}{\partial \beta}=0$. Thus, $\frac{\partial E(\pi_{R,\Gamma})}{\partial \beta}>0$ if $\beta<\hat{\beta}$ and $ \frac{\partial E(\pi_{R,\Gamma})}{\partial \beta}<0$ if $\beta>\hat{\beta}$. 

(ii) Given sellers' bids $b_H$ and $b_L$ and prices $p_A^{*}$ and  $p_B^{*}$ from Lemma \ref{biddingstrategy}, in the Independent Setup, the advertising platform's revenue is the same as the the marketplace's product listing fee in the Integrated Setup,  $E(\pi_{O,\Delta})=E(\pi_{M,\Gamma})=\frac{(1-r) L_1 L_3\alpha (\alpha -\delta)}{t (3 \alpha  \delta-2 \alpha  (1-\beta )-2 (\beta +1) \delta)^2}$

From (i) of the proof, we show $L_3>0$ given $0<\delta<\alpha<1,0 \le \beta\le 1, t>0, t \left(\frac{\alpha }{2 \delta-\alpha \delta}+\frac{1}{2}\right)<v$.  

$\frac{\partial L_3}{\partial \beta}=-2 t (\alpha  (1-\delta)-\beta  (\alpha -\delta))-t (\alpha +\delta)-v (2 \beta  (\alpha -\delta)-(2-\alpha ) \delta)<0$, given $0<\delta<\alpha<1,0 \le \beta\le 1, t>0,v>0$. 

We also showed that $L_1>0$ and $L_1$ decreases when $\beta$ increases in Proof of Lemma \ref{betaonbids}. Hence, it is easy to see that the numerator of $E(\pi_{O,\Delta})$ decreases when $\beta$ increases, since all terms are positive; and $L_1$ and $L_3$ decrease when $\beta$ increase, whereas the $(1-r)\alpha(\alpha-\delta)$ does not change. 

We showed that $3 \alpha  \delta-2 \alpha  (1-\beta )-2 (\beta +1) \delta<0$ and decreases when $\beta$ increases, given  $0<\delta<\alpha<1,0 \le \beta\le 1$. Thus, the denominator of $E(\pi_{O,\Delta})$ increases when $\beta$ increases. 

Therefore, $E(\pi_{O,\Delta})$  decreases when $\beta$ increases, and the optimal precision is the lower limit of $\beta$, i.e., $\beta^*_\Delta=0$. $\blacksquare$

\textbf{Proof of Corollary  \ref{incentive}}
From the Proof of Lemma \ref{effectofbeta}, the optimal precision $\beta^*_\Gamma=0$ if $r \le r_L$ and $\beta^*_\Gamma>0$ if $r>r_l$ in the Integrated Setup; whereas the optimal precision $\beta^*_\Delta=0$. Therefore, $\beta^*_\Gamma \ge \beta^*_\Delta$. $\blacksquare$

\textbf{Proof of Proposition \ref{twocasecompare}}\\
(i) We calculate A's expected profit from product sale as $E(\pi_{A,C})= (1-r) p_A^*(\int_{0}^{1/2} d_A^A ds+ \int_{1/2}^{1} d_A^B ds)=\frac{(1-r) L_1^2(\alpha  (1-\beta )+\delta (1-\alpha +\beta))}{2 t (3 \alpha  \delta-2 \alpha  (1-\beta )-2 (\beta +1) \delta)^2}$, and its expected adverting fee as\\
$E(\pi_{A,F})=\alpha\left(\int_{0}^{z_B^*} b_L ds+\int_{z_B^*}^{1/2} b_H ds\right)=\frac{(1-r) L_1 L_3\alpha (\alpha -\delta)}{2 t (3 \alpha  \delta-2 \alpha  (1-\beta )-2 (\beta +1) \delta)^2}$.\\
The expected profit of A is $E(\pi_{A})=E(\pi_{A,C})-E(\pi_{A,F})$, and symmetrically, $E(\pi_B)=E(\pi_A)$. 

From Proof of Lemma \ref{effectofbeta} (ii),  $E(\pi_{O,\Delta})=E(\pi_{M,\Gamma})$ and we showed $E(\pi_{O,\Delta})$ (also $E(\pi_{M,\Gamma})$) decreases when $\beta$ increases,  given  $0<\delta<\alpha<1,0 \le \beta\le 1$. 
It is easy to see that $E(\pi_{M,\Gamma})=E(\pi_{A,F})+E(\pi_{B,F})=2 E(\pi_{A,F})$. Thus, $E(\pi_{A,F})$ decreases when $\beta$ increases,  given  $0<\delta<\alpha<1,0 \le \beta\le 1$. 

$\frac{\partial E(\pi_{A,C})}{\partial \beta}= \frac{(1-r) (\alpha -\delta) (\alpha  \delta t-\beta  (\alpha -\delta) (v-t)-2 \alpha  \delta v+\delta v+\alpha  v) Z_1}{2 t (2 \alpha  (1-\beta )-3 \alpha  \delta+2 (\beta +1) \delta)^3}$, where \\$Z_1=t \left(2 \alpha ^2 (1-\beta ) (2-\beta )+\delta^2 \left(5 \alpha ^2-\alpha  (7 \beta +8)+2 (\beta +1) (\beta +2)\right)+\alpha  \delta \left(8-\alpha  (8-7 \beta )-4 \beta ^2\right)\right)$\\$-v \left(2 \alpha ^2 (1-\beta )^2+d^2 \left(4 \alpha ^2-5 \alpha  (\beta +1)+2 (\beta +1)^2\right)+\alpha  \delta \left(5 \alpha  \beta -5 \alpha -4 \beta ^2+4\right)\right)$.\\
We proved previously that $2 \alpha  (1-\beta )-3 \alpha  \delta+2 (\beta +1) \delta>0$ given  $0<\delta<\alpha<1,0 \le \beta\le 1$. Thus, the denominator of $\frac{\partial E(\pi_{A,C})}{\partial \beta}$ is positive. 

It is easy to see that $(1-r) (\alpha -\delta)>0$ and we can show that $(\alpha  \delta t-\beta  (\alpha -\delta) (v-t)-2 \alpha  \delta v+\delta v+\alpha  v)>0$, given $0<\delta<\alpha<1,0 \le \beta\le 1$, and $\frac{\alpha  \delta t-2 \delta t-2 \alpha  t}{2 \alpha  \delta-4 \delta}<v<2t$. 

We take derivative of $Z_1$ w.r.t. $v$, and get $-2 \alpha ^2 (1-\beta )^2-\delta^2 \left(4 \alpha ^2-5 \alpha  (\beta +1)+2 (\beta +1)^2\right)-\alpha  \delta \left(5 \alpha  \beta -5 \alpha -4 \beta ^2+4\right)<0$, given  $0<\delta<\alpha<1$ and $0 \le \beta\le 1$. This means $Z_1$ decreases with $v$. We evaluate $Z_1$ at the maximum of $v$, $v=2t$, we get \\$Z_1|_{v=2t}=t \left(2 \alpha ^2 (1-\beta) \beta +\delta^2 \left(\alpha  (3 \beta +2)-3 \alpha ^2-2 \beta  (\beta +1)\right)+\alpha  \delta \left(\alpha  (2-3 \beta )+4 \beta ^2\right)\right)>0$ given $0<\delta<\alpha<1,0 \le \beta\le 1$.  This means $Z_1>0$, and thereby, the numerator of $\frac{\partial E( \pi_{A,C})}{\partial \beta}$ is positive. Therefore,  $\frac{\partial E( \pi_{A,C})}{\partial \beta}>0$. \\
Since $E( \pi_{A,C})$ increases but $E(\pi_{A,F})$ decreases with $\beta$, the sellers' expected profit, $E(\pi_{A})$ and $E(\pi_{B})$, increases with $\beta$. 

It is easy to see $E(\pi_{A,\Gamma})=E(\pi_{A,\Delta})=E(\pi_{A})$, given $\beta$ . Since  $\beta^*_\Gamma\ge \beta^*_\Delta$ from Corollary \ref{incentive}, and also $E(\pi_{A})$ increases when $\beta$ increases, $E(\pi_{A,\Gamma})\ge E(\pi_{A,\Delta})$ and $E(\pi_{B,\Gamma})\ge E(\pi_{B,\Delta})$.

(ii) We formulate the consumer surplus in the case where $A$ is placed in the more-prominent position and $B$ in the less-prominent position for a consumer whose signal is  $0<s \leq 1/2$ and $B$ is placed in the more-prominent position and $A$ in the less-prominent position if $1/2 < s<1$, depending on  $p_A$ and  $p_B$.

Consumer surplus given that $0\le s\le z_B$:\\
$CS_1=(1-\beta ) z_B (\alpha \delta \left(\int_0^{1/2} (v-p_A-t s) ds+\int_{1/2}^1 (v-p_B-t (1-s)) ds\right)+\alpha(1-\delta) \int_0^{z_A} (v-p_A-t s)  ds+(1-\alpha)\delta \int_{z_B}^{1} (v-p_B-t (1-s))  ds)+\beta \alpha  \int_0^{z_B} (v-p_A-t s) ds$.\\
Consumer surplus given $z_B < s\le 1/2$:\\
$CS_2=(1-\beta ) z_B (\alpha \delta \left(\int_0^{1/2} (v-p_A-t s) ds+\int_{1/2}^1 (v-p_B-t (1-s)) ds\right)+\alpha(1-\delta) \int_0^{z_A} (v-p_A-t s)  ds+(1-\alpha)\delta \int_{z_B}^{1} (v-p_B-t (1-s))  ds)+\beta \alpha \int_{z_B}^{1/2} (v-p_A-t s) ds+\beta \delta(1-\alpha) \int_{z_B}^{1/2} (v-p_B-t (1-s)) ds$

The two sellers' strategies and listing outcomes are symmetric, which means that consumers surplus given $1/2< s \le1$ is the same as the consumer surplus given $0 \le s\le 1/2$. Thus, consumer surplus is \\$CS=2 (CS_1+CS_2)=\frac{\beta  \left((1-\alpha) \delta (2 p^*+t-2 v)^2+\alpha  t (4 v-4 p^*-t)\right)+(1-\beta) \left(2 \delta (p^*-v)^2+2 \alpha  (p^*-v)^2-\alpha  \delta (2 p^*+t-2 v)^2\right)}{4 t}$, where $p^*=p_A^*=p_B^*$ from Lemma 1.

$\frac{\partial^2 CS}{\partial \beta^2}=-\frac{(\alpha -\delta)^2 (2 \delta t+2 \alpha  t-\alpha  \delta (t+v)) \left(\delta^2 (6 (\beta +1) t -\alpha  t (11-6 \alpha +5 \beta )-\alpha  (\beta +1) v)+\alpha  \delta (t (5 \alpha  \beta -11 \alpha +12)-\alpha  (1-\beta ) v)+6 \alpha ^2 (1-\beta ) t\right)}{t (3 \alpha  \delta-2 \alpha  (1-\beta )-2 (\beta +1) \delta)^4}$. Given $0<\delta<\alpha<1$ and $0<v<2t$, we can show that $ 2 \delta t+2 \alpha  t-\alpha  \delta (t+v)>0$ and $\delta^2 (6 (\beta +1) t -\alpha  t (11-6 \alpha +5 \beta )-\alpha  (\beta +1) v)+\alpha  \delta (t (5 \alpha  \beta -11 \alpha +12)-\alpha  (1-\beta ) v)+6 \alpha ^2 (1-\beta ) t>0$. 
It is obvious other terms in $\frac{\partial^2 CS}{\partial \beta^2}$ are also positive given $0<\delta<\alpha<1$ and $0<v<2t$. Thus, $\frac{\partial^2 CS}{\partial \beta^2}<0$ (notice the negative sigh in the front). This means that $\frac{\partial CS}{\partial \beta}$ decreases with $\beta$ given $0 \le \beta \le 1$. 

$\frac{\partial CS}{\partial \beta}=-\frac{(\alpha-\delta)N_4}{4 t (2 \alpha+(2-3 \alpha) \delta )^3}$, given $\beta=0$ (the lower bound of $\beta$), where \\
$N_4=\left(\substack{(8-\alpha  (\alpha  (2-7 \alpha )+12)) \delta^3 t^2+2 \alpha  (12-\alpha  (\alpha +12)) \delta^2 t^2-v \left(4 (1-\alpha ) (2-5 (1-\alpha ) \alpha ) \delta^3 t+8 \alpha  (3-\alpha  (7-5 \alpha )) \delta^2 t+4 \alpha ^2 (6-7 \alpha ) \delta t+8 \alpha ^3 t\right)+\\v^2 \left(4 \alpha ^3+2 (1-\alpha ) (2-7 (1-\alpha ) \alpha ) \delta^3+4 \alpha  (3-\alpha  (9-7 \alpha )) \delta^2+6 \alpha ^2 (2-3 \alpha ) \delta\right)+12 (2-\alpha ) \alpha ^2 \delta t^2+8 \alpha ^3 t^2}\right)$ 

Given $0<\delta<\alpha<1$ and $0<v<2t$, we can show that $4 t (2 \alpha+(2-3 \alpha) \delta )^3>0$ and $N_4>0$. Thus,   $\frac{\partial CS}{\partial \beta}<0$, given $\beta=0$. Since $\frac{\partial CS}{\partial \beta}$ decreases when $\beta$ increases,  $\frac{\partial CS}{\partial \beta}<0$, given $0 \le\beta \le 1$.  

It is easy to see $CS_{\Gamma}=CS_{\Delta}=CS$, given $\beta$ .  Since $\beta^*_\Gamma \ge\beta^*_\Delta$ from Corollary \ref{incentive}, and also $CS$ decreases when $\beta$ increases, $CS_\Gamma \le CS_\Delta$.

(iii) We formulate the social welfare, which is the expected consumer utility minus the misfit cost of the purchased product, in the case where $A$ is placed in the more-prominent position and $B$ in the less-prominent position for a consumer whose signal is  $0<s \leq 1/2$ and $B$ is placed in the more-prominent position and $A$ in the less-prominent position if $1/2 < s<1$, depending on  $p_A$ and  $p_B$.

Social welfare given that $0\le s\le z_B$:\\
$SW_1=(1-\beta ) z_B (\alpha \delta \left(\int_0^{1/2} (v-t s) ds+\int_{1/2}^1 (v-t (1-s)) ds\right)$\\$+\alpha(1-\delta) \int_0^{z_A} (v-t s)  ds+(1-\alpha)\delta \int_{z_B}^{1} (v-t (1-s))  ds)$\\$+\beta \alpha  \int_0^{z_B} (v-t s) ds$.\\
Social welfare given $z_B < s\le 1/2$:\\
$SW_2=(1-\beta ) z_B (\alpha \delta \left(\int_0^{1/2} (v-t s) ds+\int_{1/2}^1 (v-t (1-s)) ds\right)$\\$+\alpha(1-\delta) \int_0^{z_A} (v-t s)  ds+(1-\alpha)\delta \int_{z_B}^{1} (v-t (1-s))  ds)$\\$+\beta \alpha \int_{z_B}^{1/2} (v-t s) ds+\beta \delta(1-\alpha) \int_{z_B}^{1/2} (v-t (1-s)) ds$

The two sellers' strategies and listing outcomes are symmetric, which means that social welfare given $1/2< s \le1$ is the same as social welfare given $0 \le s\le 1/2$. Thus, social welfare is 
$SW(\beta)=2 (SW_1+SW_2)=$\\$\frac{\beta  \left((1-\alpha ) \delta \left((t-2 v)^2-4 (p^*)^2\right)-\alpha  t (t-4 v)\right)+(1-\beta ) \left(2 \alpha  (v-(p^*)) ((p^*)+v)+\delta \left(2 v^2-(2-4 \alpha ) (p^*)^2-\alpha  (t-2 v)^2\right)\right)}{4 t}$, where $p^*=p_A^*=p_B^*$ from Lemma 1.

Taking the first derivative of $SW$  w.r.t. $\beta$, we get \\$\frac{\partial SW}{\partial \beta}=\frac{2 (\alpha -\delta) \left(6 t v-t^2-3 v^2\right)}{16 t}+\frac{4 \alpha  \delta (\alpha -\delta) (2 \delta t+2 \alpha  t-\alpha  \delta (t+v))^2}{(16 t) (2 \alpha  (1-\beta )-3 \alpha  d+2 (\beta +1) \delta)^3}+\frac{2 (\alpha -\delta) ((\alpha +2) \delta t-3 \alpha  \delta v+2 \alpha  t) (\alpha  \delta (t+v)-2 t (\alpha +\delta))}{(16 t) (2 \alpha  (1-\beta )-3 \alpha  \delta+2 (\beta +1) \delta)^2}$

When $t>(3-\sqrt{6})v$, $6 t v-t^2-3 v^2>0$, given $t<\frac{2 (\alpha -2) \delta v}{(\alpha -2) \delta-2 \alpha }$, $v>0$ and $0<\delta<\alpha<1$. Thus, $\frac{2 (\alpha -\delta) \left(6 t v-t^2-3 v^2\right)}{16 t}>0$, the first term of $\frac{\partial SW}{\partial \beta}$ is positive.\\
It is evident to see that the numerator of the second term of  $\frac{\partial SW}{\partial \beta}$ is positive, given $0<\delta<\alpha<1$. Moreover, we previously showed that $2 \alpha  (1-\beta )-3 \alpha  \delta+2 (\beta +1) \delta$,  which can be rewritten as $(4-3 \alpha ) \delta-2 (1-\beta ) (\alpha -\delta)$, is positive, given  $0\le \beta \le 1$ and $0<\delta<\alpha<1$. Thus, the denominator of the second term of $\frac{\partial SW}{\partial \beta}$ is positive, and thereby, the second term of  $\frac{\partial SW}{\partial \beta}$ is positive.

We now check the sign of the third term of  $\frac{\partial SW}{\partial \beta}$, $\frac{2 (\alpha -\delta) ((\alpha +2) \delta t-3 \alpha  \delta v+2 \alpha  t) (\alpha  \delta (t+v)-2 t (\alpha +\delta))}{(16 t) (2 \alpha  (1-\beta )-3 \alpha  \delta+2 (\beta +1) \delta)^2}$. 
First, we look at $ (\alpha  \delta (t+v)-2 t (\alpha +\delta))$, which increases with $v$ given  $0<\delta<\alpha<1$. We substitute the maximum value of $v$, $v<2t$ into $ (\alpha  \delta (t+v)-2 t (\alpha +\delta))$, and get $-2 (1-\alpha ) \delta t-\alpha  (1-\delta) t-\alpha  t<0$, given $0<\delta<\alpha<1$, and $t>0$. Since this term evaluated at the maximum value of $v$ is negative, $ (\alpha  \delta (t+v)-2 t (\alpha +\delta))<0$ given $0<\delta<\alpha<1$ and $v<2t$ (as well as $t>(3-\sqrt{6})v$). 
Then, we look at $(\alpha +2) \delta t-3 \alpha  \delta v+2 \alpha  t$. By equating it to $0$, we get $\delta>\frac{2 \alpha  t}{3 \alpha  v-\alpha  t-2 t}$ and $\alpha>\frac{4 t}{3v-t}$ so that $(\alpha +2) \delta t-3 \alpha  \delta v+2 \alpha  t<0$. Combine  $\delta>\frac{2 \alpha  t}{3 \alpha  v-\alpha  t-2 t}$, $\alpha>\frac{4 t}{3v-t}$ and   $\delta<\alpha$, we get $\delta>\frac{4 t}{3v-t}$, which ensures $(\alpha +2) \delta t-3 \alpha  \delta v+2 \alpha  t<0$, given $0<\delta<\alpha<1$ and $v<2t$ (as well as $t>(3-\sqrt{6})v$). 
Hence, when  $\delta>\frac{4 t}{3v-t}$, the two terms in the numerators of third term of $\frac{\partial SW}{\partial \beta}$, $ (\alpha  \delta (t+v)-2 t (\alpha +\delta))<0$ and $(\alpha +2) \delta t-3 \alpha  \delta v+2 \alpha  t<0$, whereas the remaining terms in the numerator and all terms in the denominator are positive, given  $0<\delta<\alpha<1$ and $0<v$; thereby,  the third term of  $\frac{\partial SW}{\partial \beta}$ is positive. 

Therefore, when $t>(3-\sqrt{6})v$ and  $\delta>\frac{4 t}{3v-t}$, all three terms of  $\frac{\partial SW}{\partial \beta}$ are positive, and thereby,  $\frac{\partial SW}{\partial \beta}$ is positive.

It is easy to see social welfare are the same in both setups, i.e., $SW_{\Delta}=SW_{\Gamma}=SW(\beta)$, given $\beta$ is the same. Since  $\beta^*_\Gamma > \beta^*_\Delta$ if $r> r_l$ from Lemma \ref{effectofbeta}, and also $SW$ increases when $\beta$ increases, $SW_\Gamma > SW_\Delta$. 

$\blacksquare$

\end{spacing}
\end{document}